\documentclass{elsart}
\usepackage{epsfig}
\usepackage{amsmath}
\usepackage{amssymb}

\begin{document}
\begin{frontmatter}
\title{Search for diffuse neutrino flux 
 from astrophysical sources with MACRO}
\begin{center}
{\rm The MACRO Collaboration} \\
\nobreak\bigskip\nobreak
\pretolerance=10000
M.~Ambrosio$^{12}$, 
R.~Antolini$^{7}$, 
G.~Auriemma$^{14,a}$, 
D.~Bakari$^{2,17}$, 
A.~Baldini$^{13}$, 
G.~C.~Barbarino$^{12}$, 
B.~C.~Barish$^{4}$, 
G.~Battistoni$^{6,b}$, 
Y.~Becherini$^{2}$,
R.~Bellotti$^{1}$
% ...
\\
\vspace{1.5 cm}
\footnotesize
1. Dipartimento di Fisica dell'Universit\`a  di Bari and INFN, 70126 Bari, Italy 
\\
2. Dipartimento di Fisica dell'Universit\`a  di Bologna and INFN, 40126 Bologna, 
Italy \\
3. Physics Department, Boston University, Boston, MA 02215, USA \\
4. California Institute of Technology, Pasadena, CA 91125, USA \\
5. Department of Physics, Drexel University, Philadelphia, PA 19104, USA \\
6. Laboratori Nazionali di Frascati dell'INFN, 00044 Frascati (Roma), Italy \\
7. Laboratori Nazionali del Gran Sasso dell'INFN, 67010 Assergi (L'Aquila), 
Italy \\
8. Depts. of Physics and of Astronomy, Indiana University, Bloomington, IN 
47405, USA \\
9. Dipartimento di Fisica dell'Universit\`a  dell'Aquila and INFN, 67100 
L'Aquila, Italy\\
10. Dipartimento di Fisica dell'Universit\`a  di Lecce and INFN, 73100 Lecce, 
Italy \\
11. Department of Physics, University of Michigan, Ann Arbor, MI 48109, USA \\
12. Dipartimento di Fisica dell'Universit\`a  di Napoli and INFN, 80125 Napoli, 
Italy \\
13. Dipartimento di Fisica dell'Universit\`a  di Pisa and INFN, 56010 Pisa, 
Italy \\
14. Dipartimento di Fisica dell'Universit\`a  di Roma "La Sapienza" and INFN, 
00185 Roma, Italy \\
15. Physics Department, Texas A\&M University, College Station, TX 77843, USA \\
16. Dipartimento di Fisica Sperimentale dell'Universit\`a  di Torino and INFN, 
10125 Torino, Italy \\
17. L.P.T.P, Faculty of Sciences, University Mohamed I, B.P. 524 Oujda, Morocco 
\\
18. Dipartimento di Fisica dell'Universit\`a  di Roma Tre and INFN Sezione Roma 
Tre, 00146 Roma, Italy \\
$a$ Also Universit\`a  della Basilicata, 85100 Potenza, Italy \\
$b$ Also INFN Milano, 20133 Milano, Italy \\
$c$ Also Istituto IASF/CNR, 40129 Bologna, Italy \\
$d$ Also Universit\`a  di Trieste and INFN, 34100 Trieste, Italy \\
$e$ Also Dipartimento di Energetica, Universit\`a  di Roma, 00185 Roma, Italy \\
$g$ Also Institute for Space Sciences, 76900 Bucharest, Romania \\
$h$ Macalester College, Dept. of Physics and Astr., St. Paul, MN 55105 \\
$i$ Also Department of Physics, Pennsylvania State University, University Park, 
PA 16801, USA \\
$l$ Also Dipartimento di Fisica dell'Universit\`a  della Calabria, Rende 
(Cosenza), Italy \\
$o$ Also Dipartimento di Scienze e Tecnologie Avanzate, Universit\`a  del 
Piemonte Orientale, Alessandria, Italy \\
$p$ Also Duluth Physics Department, University of Minnesota, Duluth, MN 55812 \\
$q$ Also Dept. of Physics, MIT, Cambridge, MA 02139 \\
$r$ Also Intervideo Inc., Torrance CA 90505 USA \\
$s$ Also Resonance Photonics, Markham, Ontario, Canada\\
$ $\\
$ $\\
* Corresponding author. Email: lorenzo.perrone@le.infn.it 
\end{center}
\newpage
\begin{abstract}
Many galactic and extragalactic astrophysical sources are currently considered 
promising candidates as high energy neutrino emitters. 
Astrophysical neutrinos can be detected as upward-going muons produced in 
charged-current interactions with the medium surrounding the detector. The 
expected neutrino fluxes from various models start to dominate on the 
atmospheric neutrino background at neutrino energies above some tens of TeV.            
We present the results of a search for an excess of high energy upward-going 
muons among the sample of data collected by MACRO during $\sim~5.8$ years of 
effective running time.  No significant evidence for this signal was found. As a 
consequence, an upper limit on the flux of upward-going muons from high-energy 
neutrinos was set at the level of $1.7$~$\times 10^{-14}  \mathrm{cm^{-2}} \, 
\mathrm{s^{-1}} \, \mathrm{sr^{-1}} $.  The corresponding upper limit for the 
diffuse neutrino flux was evaluated assuming a neutrino power law spectrum. Our 
result was compared with theoretical predictions and upper limits from other 
experiments.  
\end{abstract}
\end{frontmatter}
\section{Neutrino astronomy: overview and motivation}
\vspace{-0.2cm}
Neutrinos with energies larger than $\sim 1$ GeV are expected from a wide class 
of galactic and extragalactic astrophysical sources. Neutrino production 
requires the existence of hadronic processes and it is generally described in 
the picture of the {\it beam dump model}~\cite{gp}: high  energy protons 
accelerated  close to  compact objects  by shock waves or plasma turbulence 
interact with photons or target matter surrounding the source, producing pions.
  Neutrinos of electron and muon flavors  originate from  decays of charged 
pions, as well from  subsequent muon decays.  In the same hadronic chains, high 
energy $\gamma$-rays are expected to be produced through neutral pion decays. 
Like $\gamma$-rays, neutrinos can travel undeflected through the Universe. 
Neutrinos however are much less absorbed than photons and thus make a more 
powerful {\em probe} for astronomy searches.
Many of the candidate sources of neutrinos  (binary systems, supernovae 
remnants, AGNs, GRBs, etc) have already been recognized as gamma rays emitters  
  at energies higher than 1 TeV: this provides an important hint for neutrino 
astronomy, even though the observed $\gamma$-ray energies are not high enough to 
exclude the electromagnetic production mechanisms, such as synchrotron or 
inverse Compton processes. In this scenario, the detection 
of high energy neutrinos  would open a new field of research, complementary to 
$\gamma$-ray  astronomy. 
 
Neutrinos coming from a source propagate through the Earth, occasionally 
producing upward-going muons by charged-current interactions with the matter 
surrounding the detector. The detection probability grows with the energy due to 
the increase of both neutrino-nucleon cross sections and muon range, so that 
the effective detector mass results from the convolution
of the detector area with the muon path-length. 

Underwater/ice neutrino detectors like Baikal~\cite{baikal} and 
AMANDA~\cite{amanda}, as well as  
underground experiments like MACRO~\cite{macroastro}, IMB~\cite{imb}, 
Baksan~\cite{Baksan} and Superkamiokande~\cite{SKastro}, surveyed  the sky in 
order to search for point source of astrophysical neutrinos. 
At the present their effective area is not large enough to measure the expected 
flux but they set experimental upper limits that guide and constrain 
theoretical models.
Next generation neutrino telescopes 
like ANTARES~\cite{webantares}, 
Icecube~\cite{webicecube}, NEMO~\cite{webnemo} and NESTOR~\cite{webnestor}, with effective areas ranging from $ 0.1$ to 1 Km$^{2}$, are 
expected to provide the sensitivity required for observing astrophysical 
sources. A detailed review dealing with physics goals and detection techniques 
in the field of neutrino astronomy can be found in~\cite{review2000}. 

The results of a search for point-like sources was recently presented by the 
MACRO collaboration in~\cite{macroastro}. Here we show the results of a search 
for a diffuse neutrino flux from unresolved sources, by looking for an excess of 
high energy upward-going events. A preliminary study of the sensitivity of the 
MACRO detector to high-energy muons is in~\cite{corona}.
In sec.~\ref{macrosci} we discuss the capability of the MACRO detector as a 
neutrino telescope. 
In particular, since most of events induced by  astrophysical neutrinos
 are expected in the 0.1-100 TeV energy range, the response of 
scintillation counters was accurately studied for high energy events, using the 
scintillator calibrations.
In sec~\ref{analysis} we show that the energy released in the scintillation 
counters by crossing muons is the most important information for the selection 
of events induced by  astrophysical neutrinos. 
We describe the analysis cuts on the events from a 
simulated signal, and in sec.~\ref{sec::4} on the events from the expected background. The 
main background source are upward-going muons induced from atmospheric 
neutrinos. The GEANT-based~\cite{geant} simulation tool used by 
MACRO~\cite{techpap} was modified to correctly handling the propagation of very 
high energy muons in the rock surrounding the detector~\cite{geant321}, and the 
detector response for events with high energy releases.
In sec.~\ref{sec::5} we present the results of the analysis of the real data sample, and we 
give the upper limit for the diffuse neutrino flux from unresolved sources. This 
upper limit is compared with some theoretical predictions and with upper limits 
from other experiments.  
%
\section{MACRO as a neutrino telescope}  
\label{macrosci} 
MACRO, located at the Gran Sasso Laboratories (Italy), was a large area 
multipurpose underground detector, in the shape of  a rectangular box whose 
global dimensions were  
  $76.6 \times 12.0\times 9.3$ m$^3$~\cite{techpap}.
The lower half of the apparatus was filled with rock absorber 
 alternating with streamer tube planes for particle tracking. 
  Liquid scintillator layers,    
  placed at the bottom, the center, the top and all around the 
   detector provided time information for discriminating the direction of 
    incoming particles.
  The horizontal streamer tube planes were equipped with pick-up strips 
   providing stereo read-out of the detector hits. 
 The large acceptance ($\sim$ 10$^4$ m$^2$ sr for an isotropic flux), 
  the good shielding of the site (the rate of cosmic ray muons 
  is $\sim$ 10$^{-6}$ times   
    the surface rate), the fast timing (time resolution $\sim 0.5$ ns) and 
    the good pointing capability (intrinsic angular resolution $\lesssim 1^o$)
   made MACRO suitable for working as 
   a neutrino {\it telescope}. 
%
\subsection{Scintillator response and photomultiplier saturation}
The energy deposited in a MACRO scintillation counter by a throughgoing particle 
is reconstructed by the ERP (Energy Reconstruction Processor) system according 
to the procedure described in~\cite{techpap}. 

The ERP circuit uses a twelve-bit ADC that makes an accurate integration of the 
photomultiplier pulse. This is applied to both the raw signal and a 10$\times$ 
attenuated version; the latter was used when the non attenuated ADC saturates, 
i.e. for fast magnetic monopole searches~\cite{macrosatura} and in this 
analysis. The ERP provides also time of flight from two TDCs readout.
The ERP ADCs and TDCs were calibrated weekly with delayed LED signals and 
variable-intensity laser pulses. The ADCs energy calibrations assume a linear 
relationship between the number of attenuated ADC counts and the light hitting 
the phototube $L$:
\begin{equation}     
L=G(ADC-p) 
\label{erplinear}
\end{equation}
where $G$ and $p$ are the gain factor and the pedestal of each scintillation 
counter, respectively. This hypothesis is valid for most of the analyses based on MACRO data.

The present analysis deals with signals greatly exceeding the average energy 
released in the detector by downward-going cosmic ray muons and atmospheric 
neutrino-induced events. If the energy released in a counter exceeds 4$\div$5 times the 
level of that released by minimum ionizing cosmic ray muons ($\simeq 34$ MeV 
for one vertical track crossing a horizontal scintillation counter), the 
phototube response is not linear anymore~\cite{macrosatura}. 
The effect of photomultiplier saturation can be taken into account by including 
in eq. 1 the next-to-leading order terms. By following the approach described 
in~\cite{macrosatura}, eq.~\ref{erplinear} becomes:     
\begin{equation}
L=G(ADC-p)+\mathcal{Q}(ADC-p)^2+\mathcal{T}(ADC-p)^3+ 
            \mathcal{F}(ADC-p)^4
\label{non-linear}
\end{equation} 
the constants $\mathcal{Q}$, $\mathcal{T}$, $\mathcal{F}$ represent the second-, 
the third-, and the fourth-order corrections. 
These parameters were determined (one set for each MACRO scintillation counter) 
by fitting the laser light yield $L$ versus the ERP $(ADC-p)$ value with a 
fourth degree polynomial. $G$ and $p$ are fixed at the value given by the 
standard ERP energy calibration.   

\begin{figure}[htb]
\begin{center}
\epsfig{figure=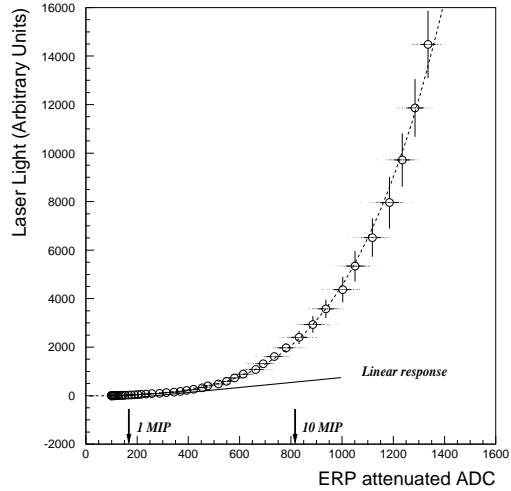, width=7.3cm,height=7.3cm}
\epsfig{figure=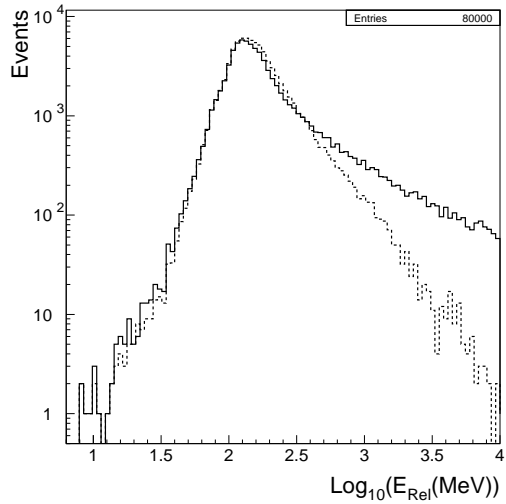,width=7.3cm,height=7.3cm}
\caption{{\it (Upper)} ERP attenuated-ADC response to laser light with linear 
and fourth-order fits. The plot is for one of the horizontal counters. The 
single ({\em 1 MIP}) and ten-times 
 ({\em 10 MIP}) minimum ionizing particle levels are shown 
 (for a vertical track crossing the center of the counter). {\it (Lower)} 
Distribution of the total released energy $E_{Rel}$ for a sample of 80000 downward-going 
muons in MACRO; dashed line: $E_{Rel}$ calculated with the linear response 
eq. 1;  solid line: $E_{Rel}$ calculated with eq. 2.}
\label{boxeps}
\end{center}
\end{figure}

In Fig.~\ref{boxeps} (upper panel), the linear and the non-linear response 
regions are shown for one of the horizontal counters.
The empty circles represent the mean values of the gaussian distributed ADC 
response for a fixed light level. The dashed curve shows the fourth-order fit 
done by taking into account the statistical errors.
The standard energy reconstruction is reliable below few times the {\em MIP} 
({\em M}inimum {\em I}onizing {\em P}article) level. This analysis searches for 
events with a large energy release, above a certain threshold. 
For this reason, we used the correction given in eq. 2, which makes the energy 
reconstruction reliable above 3 {\em MIP} and up to  
$\sim$ 15 {\em MIP} ($i.e. \ \sim 500$ MeV in a single counter). Above this 
value, the energy reconstruction may be affected by a large error because of the 
saturation. The fit constants $\mathcal{Q}$, $\mathcal{T}$, $\mathcal{F}$ and 
the parameters $G$ and $p$ of eq. 2 were supposed independent on the time.  
This approximation is verified within 10\% accuracy for the constants $G$ and 
$p$ measured with the standard calibration procedures.
For the purpose of our analysis, a reconstructed energy $\geq 500$ MeV will be 
considered as an indication of a large energy release. 

The lower panel of Fig.~\ref{boxeps} shows the distribution of the total energy 
released in the scintillators by a sample of 80000 real events collected by 
MACRO (downward-going cosmic ray muons). In particular, the solid line is obtained 
using eq. 2, whereas the dashed line is obtained  using the standard 
calibrations (eq. 1). 
The corrections act only at large values of released energy $E_{Rel}$, while 
around the distribution maximum their effect is negligible.
Ignoring the photomultiplier non-linearity leads to a systematic underestimate 
of the reconstructed energy loss for a small fraction of high energy events.      
This is in agreement with what it is expected from the curve given on the upper 
panel of Fig.~\ref{boxeps} in the saturation region: a fixed ADC value 
corresponds to a value of the reconstructed energy higher than in the case of 
the linear response approximation.

\section{Monte Carlo simulation of the AGN signal}
\label{analysis}
This analysis aims to select a sample of very high energy upward-going muons, 
since at neutrino energies above some tens of TeV the predicted neutrino fluxes 
from astrophysical sources start to rise above the atmospheric neutrino 
background. 
In our simulation of the AGN signal in MACRO we used as input    
the model by Stecker {\it et al.} \cite{Stecker} for the photo-pion production 
of neutrinos in AGN core.

Because high energy neutrino absorption from the Earth is an important effect, 
the neutrino propagation through the Earth was done by solving the kinetic 
equation for the transport of neutrinos in dense media. This approach is 
suggested in \cite{naumov}, where the neutrino absorption and the contribution 
of neutrino-nucleon  neutral-current interactions were properly taken into 
account.
The deep inelastic cross sections were calculated using the set of parton 
density functions CTEQ3-DIS~\cite{cteq}. The new version of parton density 
functions by CTEQ group~\cite{cteq5} does not produce significant changes for 
our calculation.
The neutrino-induced muon propagation in the rock outside the detector was 
evaluated using the analytical formulas given in~\cite{Lohmann,geant321}. 

Then, following the analytical distributions, we simulated a sample of 13305 
upward-going muons on the surface of a volume containing the detector plus 2 
meter-wide layer of surrounding rock. The rock was included to take into account 
the effect of electromagnetic showers induced by the muons in the detector.
In this volume, the transport of muons was done with the tool developed in 
\cite{geant321}, which correctly propagate very high energy muons up to the PeV 
energies. This software package replaced in our simulation the default transport 
modules implemented in the GEANT~\cite{geant} package.

By normalizing to the expected event rate from the model by Stecker et {\it al.} 
(4.45 events/year on the surface of this box) our simulated sample is equivalent 
to $T_{eq}=2988.5$ years.  
7547/13305 simulated muons reach the active detector and hit at least one 
scintillation counter in two different layers of the MACRO detector. 
This was the minimum requirement to define an "event".     

We adopt the following notation: taking as a reference the upper counter which 
measures the time $T_{1}$, the time of flight  $\Delta T = T_{2} - T_{1}$ is 
positive if the particle travels downward and it is negative if the particle 
travels upwards.
Two or more adjacent scintillator hits within a time window of 2 ns and on the 
same detector layer, at a maximum distance of 1 m, form a {\it scintillator 
cluster}.  
Any association between a cluster and a single hit placed on different layers, 
as well as between two different clusters on different layers defines a {\it 
scintillator track}. The {\it scintillator track} length is the geometrical 
distance between the positions of the center of each cluster. If a  {\it 
cluster} contains more than one counter, its center is calculated by averaging 
the hit positions weighted with the released energy.  
For each {\it scintillator track}, the quantity $1/\beta = c \Delta T/L$ ($L$ is 
the track length and $c$ the speed of light) in our convention is around +1 for 
downward-going particles and -1 in the opposite case.
\begin{figure}[htb]
\begin{center} 
\hskip -7.5cm
\epsfig{file=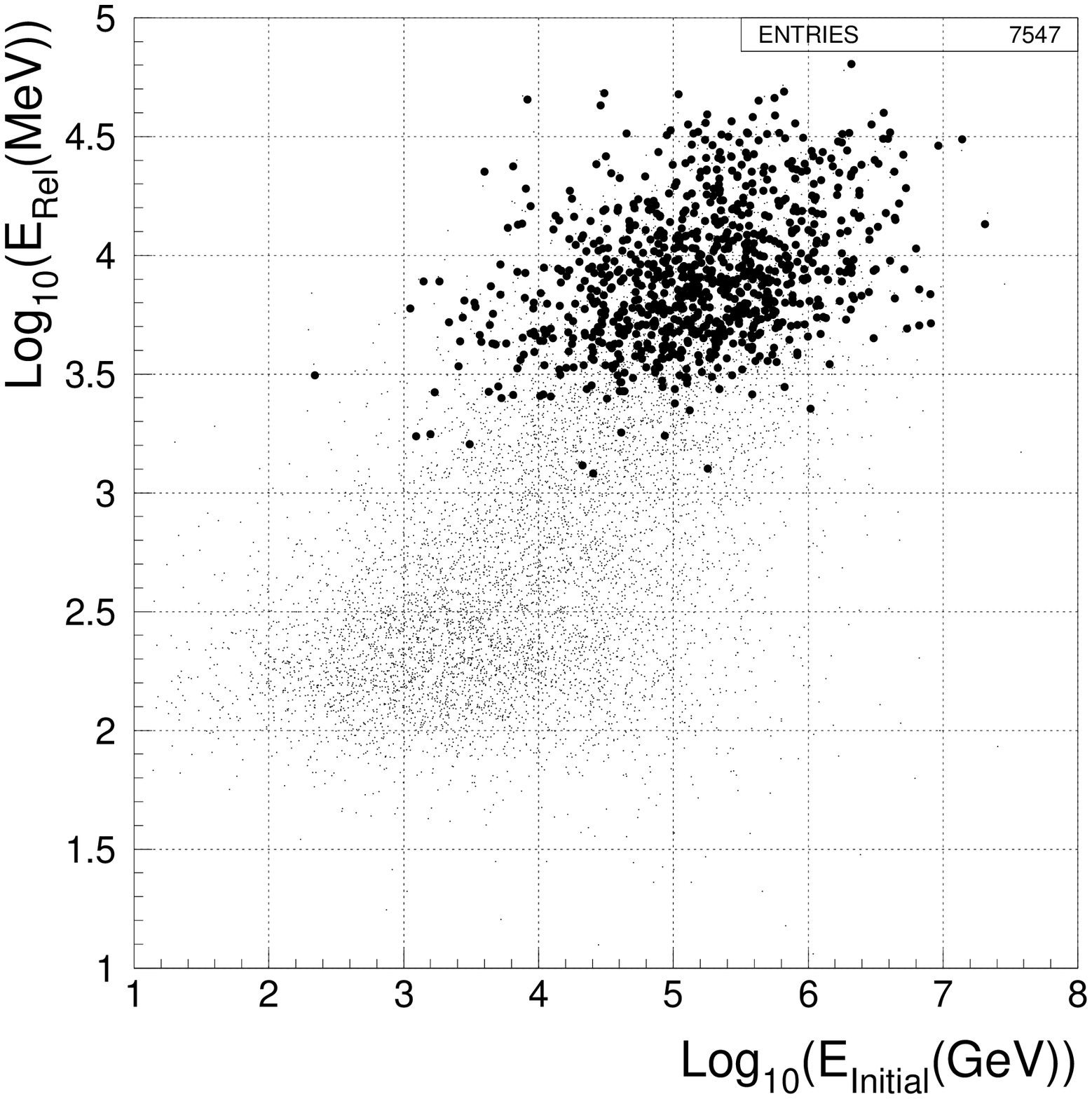,height=7.4cm,width=7.4cm}
\vskip -7.4 cm
\hskip 6.5 cm
\epsfig{file=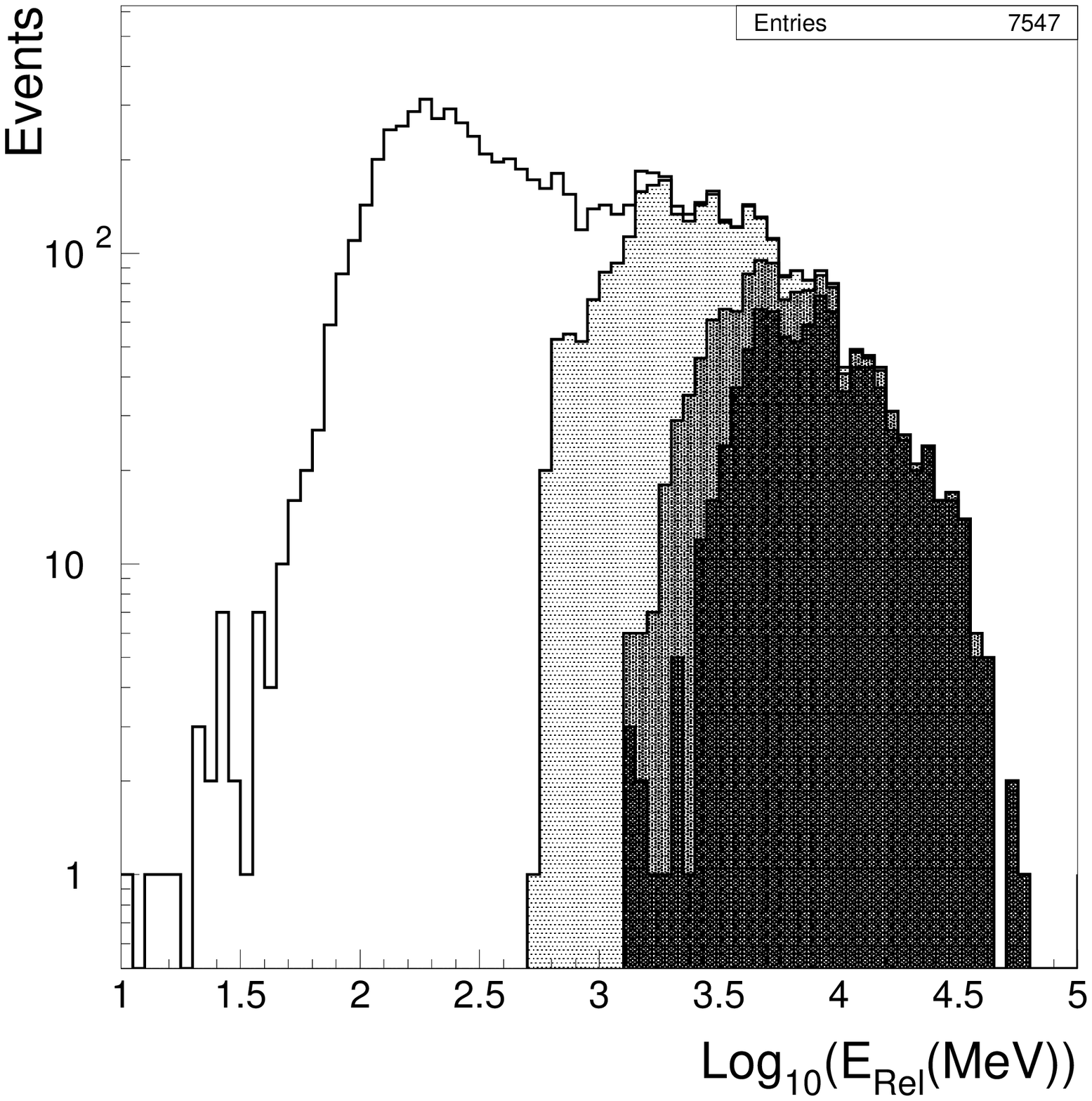,height=7.4cm,width=7.4cm}
\caption{{\it (Left) } Scatter plot of the reconstructed total energy $E_{Rel}$ 
released in the scintillator counters {\it vs.} the muon initial energy 
$E_{Initial}$. Each simulated event is represented with a point. The darker 
points represent the subsample of events which survived all the cuts described 
in the text. {\it (Right)} The effect of more and more stringent cuts (see text) 
on $E_{Rel}$ to select the subsample of high energy muons.}  
\label{corre}
\end{center}
\end{figure}

Fig.~\ref{corre} (left) shows the correlation between the reconstructed total 
energy released in the scintillator $E_{Rel}$, with the muon initial energy 
$E_{Initial}$. By initial energy we mean the energy of the muons when they enter 
the detector.
Each simulated event is represented with a point in the scatter plot. 

In order to reduce the background due to atmospheric neutrino-induced muons (see 
next section), we select the Very High Energy (VHE) muons from the AGN simulated 
signal by imposing more and more stringent conditions, the effects of which are 
illustrated in Fig.~\ref{corre}(right). 
The light shaded histogram corresponds to events for which there is at least one 
scintillator with energy release greater than 500 MeV; the histogram with the 
dark shaded area corresponds to events with at least two scintillators, each one 
with energy release greater than 500 MeV, and distance between the hits smaller 
than 1 m; the histogram with the black shaded area corresponds to events with a 
further scintillator in a different detector layer and with energy release 
greater than 500 MeV. The darker points on Fig.~\ref{corre} (left)  represent 
the subsample of events, which survived all the cuts described above. 
Hereafter we will refer to the last condition as $E_{cut}$.   
In Fig.~\ref{corre}(right) the principal maximum of the light shaded 
distribution is due to the muons which lose energy through standard ionization. 
The second ``bump'' is populated by higher energy muons, with large energy 
release in the scintillators through radiative processes.
The value of $E_{Rel}$= 500 MeV was chosen after the calibration procedure 
described in sec. 2. 
%

To discriminate the direction of the incoming events an algorithm based on the 
time-of-flight technique was developed.
The algorithm calculates the mean time associated with each scintillator layer 
involved in the event. 
We required  a reconstructed 
upward-going 
flight direction between at least two different scintillator layers.  
This condition for selecting upward-going events was called $D1_{cut}$. 

Many $scintillator\ tracks$ (associations between any two 
$scintillator\ clusters$) are present in a showering event.
Thus, a more selective cut for discriminating the upward direction, 
can be imposed 
using all the $scintillator\ tracks$ in the event. 
We required that the ratio between the number of tracks with positive $1/\beta$ 
value (downward-going) and negative $1/\beta$ (upward-going) is less than 0.4 
(cut $D2_{cut}$). The value of the ratio was optimized using a Monte Carlo study 
on a simulated sample of $\sim 10^6$ downward-going muons: no 
downward-going muons survived the energy cut $E_{cut}$ combined with the 
direction cut $D_{cut}$, defined as $D1_{cut}$ + $D2_{cut}$. After the cuts 
$E_{cut}$ + $D_{cut}$, 438/13305 simulated VHE neutrino-induced muons (signal) 
survived.

\begin{figure}[htb]
\begin{center}
\epsfig{file=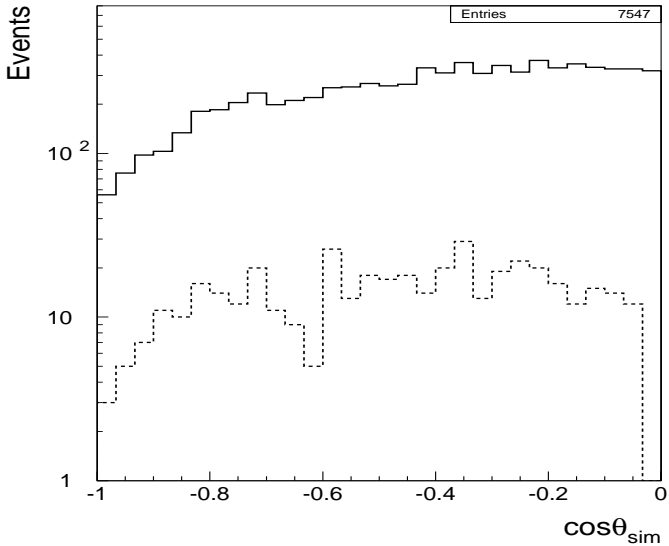,height=8cm,width=10cm}
\caption{Distribution of the cosine of zenith angle for the overall sample of 
simulated events (solid line). The dashed histogram refers to the events 
selected by the energy cut $E_{cut}$ combined with the direction cut $D_{cut}$.}  
\label{simzen}
\end{center}
\end{figure}

Fig.~\ref{simzen} shows the distribution of the simulated cosine of muon zenith 
angle $\theta_{sim}$. The solid histogram refers to the overall sample of 
simulated events; the dashed histogram represents the sample selected after the 
$E_{cut}$ + $D_{cut}$.
The $\cos \theta_{sim}$ dependence of the muon flux is a consequence of high-energy 
neutrino absorption of the Earth~\cite{quigg}. Even assuming an isotropic 
neutrino flux~\cite{Stecker} as input model, the zenith angular distribution of 
muons reaching the detector shows a strong suppression at $\cos \theta_{sim} 
\leq -0.8$ due to the hard discontinuity of the Earth density profile in 
correspondence of the massive nucleus.

\begin{figure}[htb]
\begin{center} 
\hskip -7.5cm
\epsfig{file=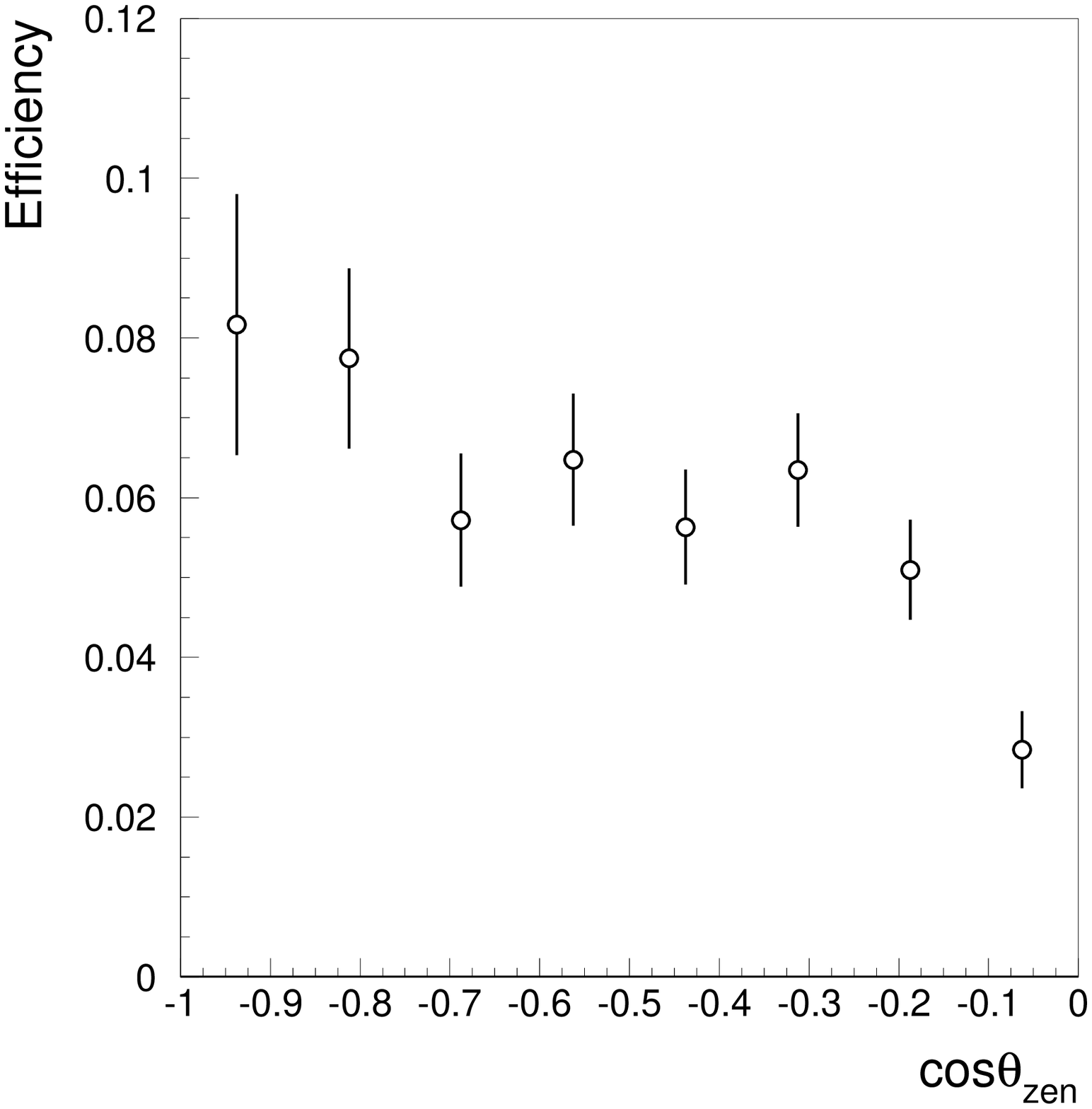,height=7.4cm,width=7.4cm}
\vskip -7.4 cm
\hskip 6.5 cm
\epsfig{file=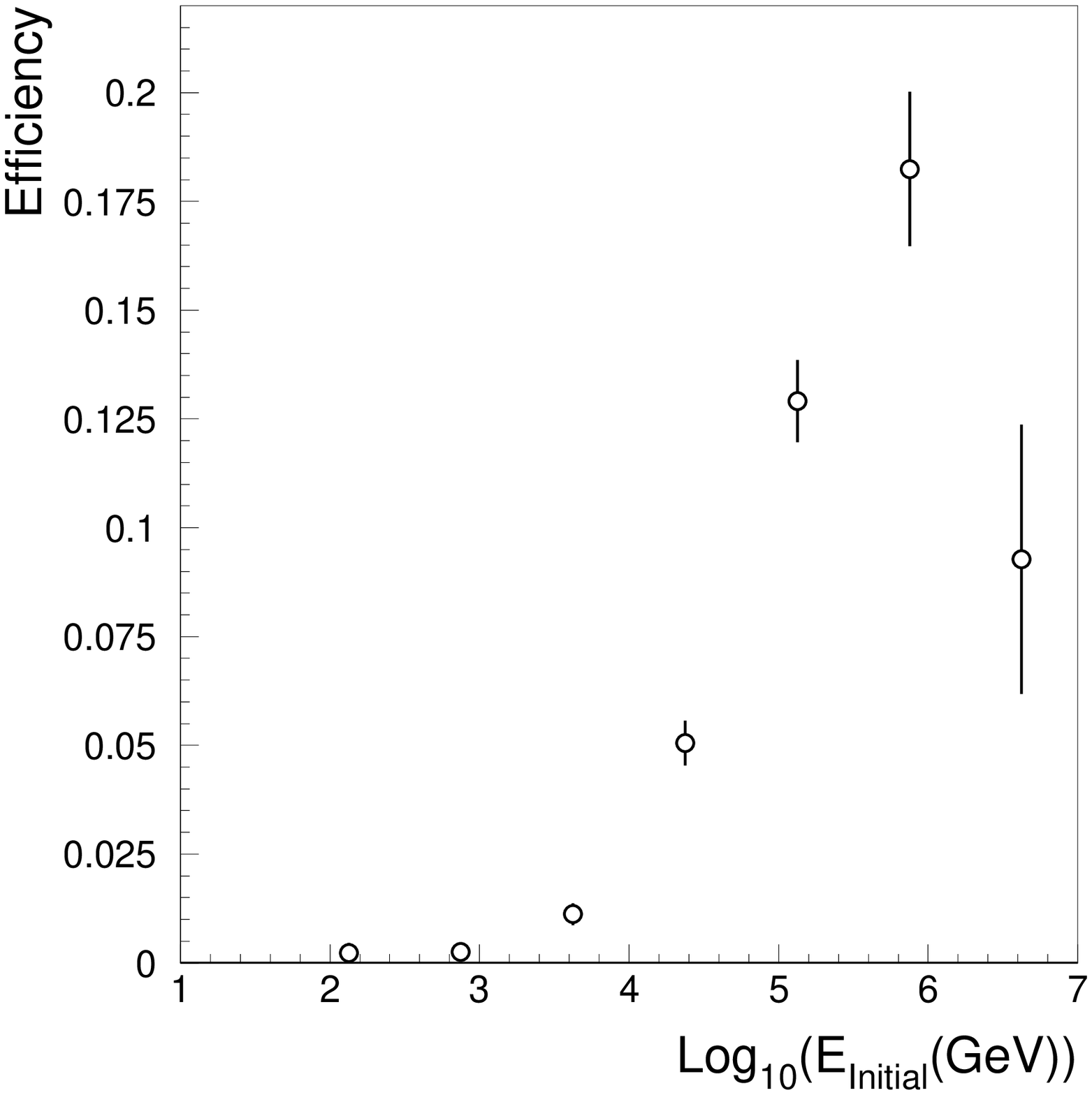,height=7.4cm,width=7.4cm}
\caption{Efficiency of analysis cuts $E_{cut}$ + $D_{cut}$  for selecting muons 
from the simulated AGN signal, as a function of the muon zenith angle {\it 
(left)} and of the muon initial energy {\it (right)}. Only the statistical 
errors due to the limited simulation sample are shown.}  
\label{effene}
\end{center}
\end{figure}

Fig.\ref{effene} shows the selection efficiency of the analysis cuts 
$E_{cut}$ + $D_{cut}$, when applied to the simulated sample of events from 
AGNs.  
In the left plot such efficiency is shown as a function of the cosine of
muon zenith angle $\theta_{zen}$,
while the right one represents the dependence on the muon initial
energy.

The energy cut $E_{cut}$ and the flight direction cut $D_{cut}$ were not 
completely efficient to reject atmospheric multiple muons and nearly horizontal 
high-energy cosmic ray muons. 
These events generate large showers in the apparatus, with a 
large number of {\it scintillator tracks}.
Thus, a final cut was defined in order to improve the 
capability of selecting upward-going events against the background of 
downward-going showering events.
For each event, we define as:      
{\it itot}, the number of scintillator tracks whose length is greater than 2.5 
m; 
{\it iup}, the number of scintillator tracks for which $-1.25 \leq 1/\beta \leq 
-0.75$;  
{\it idown}, the number of scintillator tracks for which  $0.75 \leq 1/\beta 
\leq 1.25$.
The final cut (called $S_{cut}$: $ itot > 11$ and $ iup/itot > 0.3$ and 
$idown/iup < 0.1$) selects upward-going events for events with a large number of  
$scintillator\ tracks$. This cut 
was tuned to reject Monte Carlo simulated background events and the large 
sample of multiple muons and nearly horizontal high-energy cosmic ray muons 
in the real data. 
After this last cut, 247/13305 simulated VHE signal events survived.

%
\section{The background of atmospheric neutrinos} 
\label{sec::4}
The background from atmospheric neutrinos was estimated with the Monte Carlo 
program developed for the study of the atmospheric neutrino-induced upward-going 
muon flux in MACRO.  It uses the Bartol neutrino flux~\cite{bartol}, the Morfin
and Tung parton density functions~\cite{morfin}   for the calculation of the 
DIS $\nu N$ cross sections, and the Lohmann {\it et al.}~\cite{Lohmann} muon 
energy loss for the propagation of induced muons through matter. The theoretical 
uncertainty on the expected event rate in MACRO for atmospheric neutrinos is 
$\sim$ 17\%, mainly due to the uncertainty on the neutrino flux.  Further 
details can be found in~\cite{montaruli-spurio,upmu,lowenergy}.  

\begin{figure}[tb]
\begin{center}
\hskip -7.6 cm
\epsfig{file=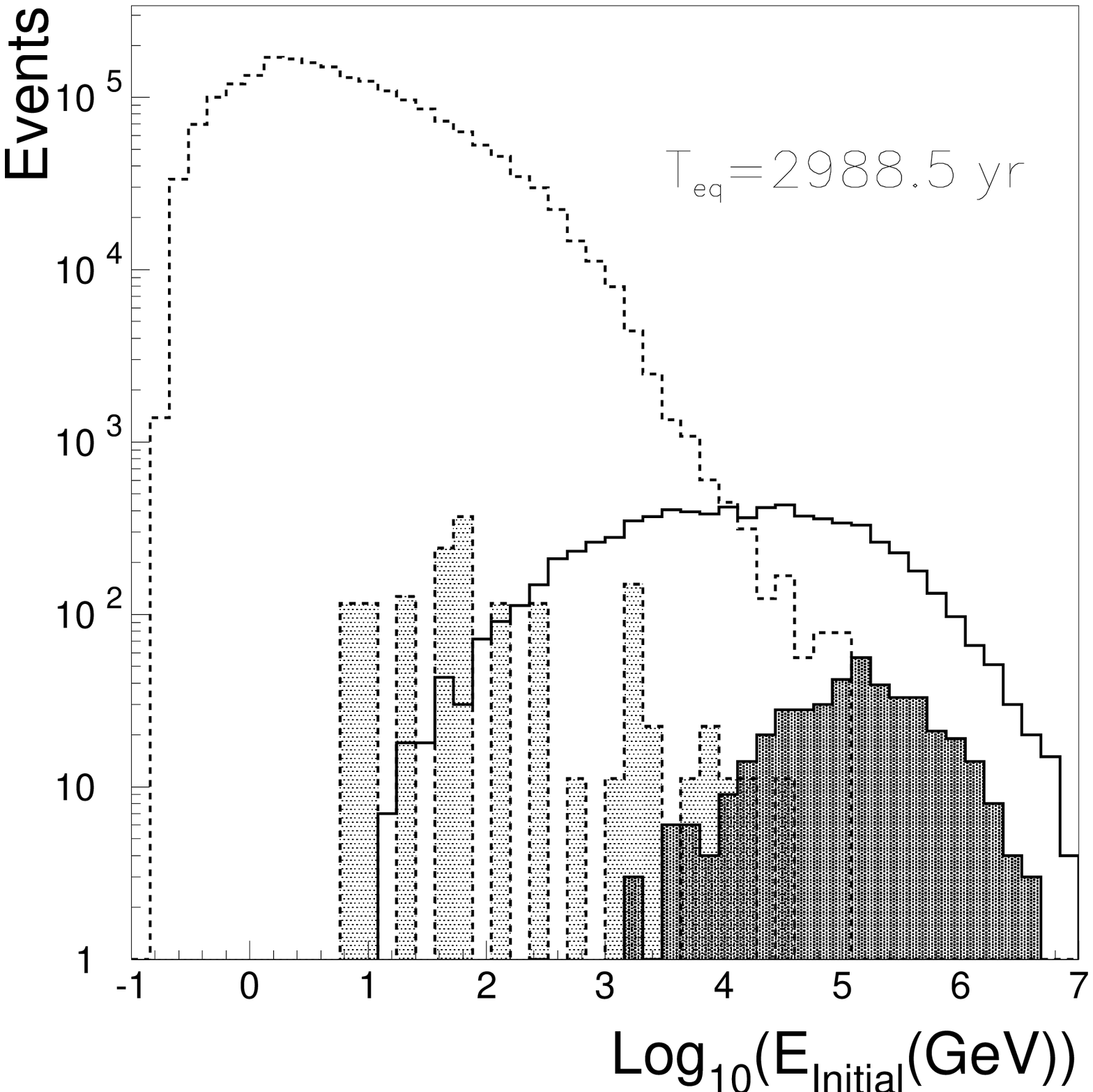,height=7.8cm,width=7.6cm}
\vskip -7.8cm
\hskip +6.3cm 
\epsfig{file=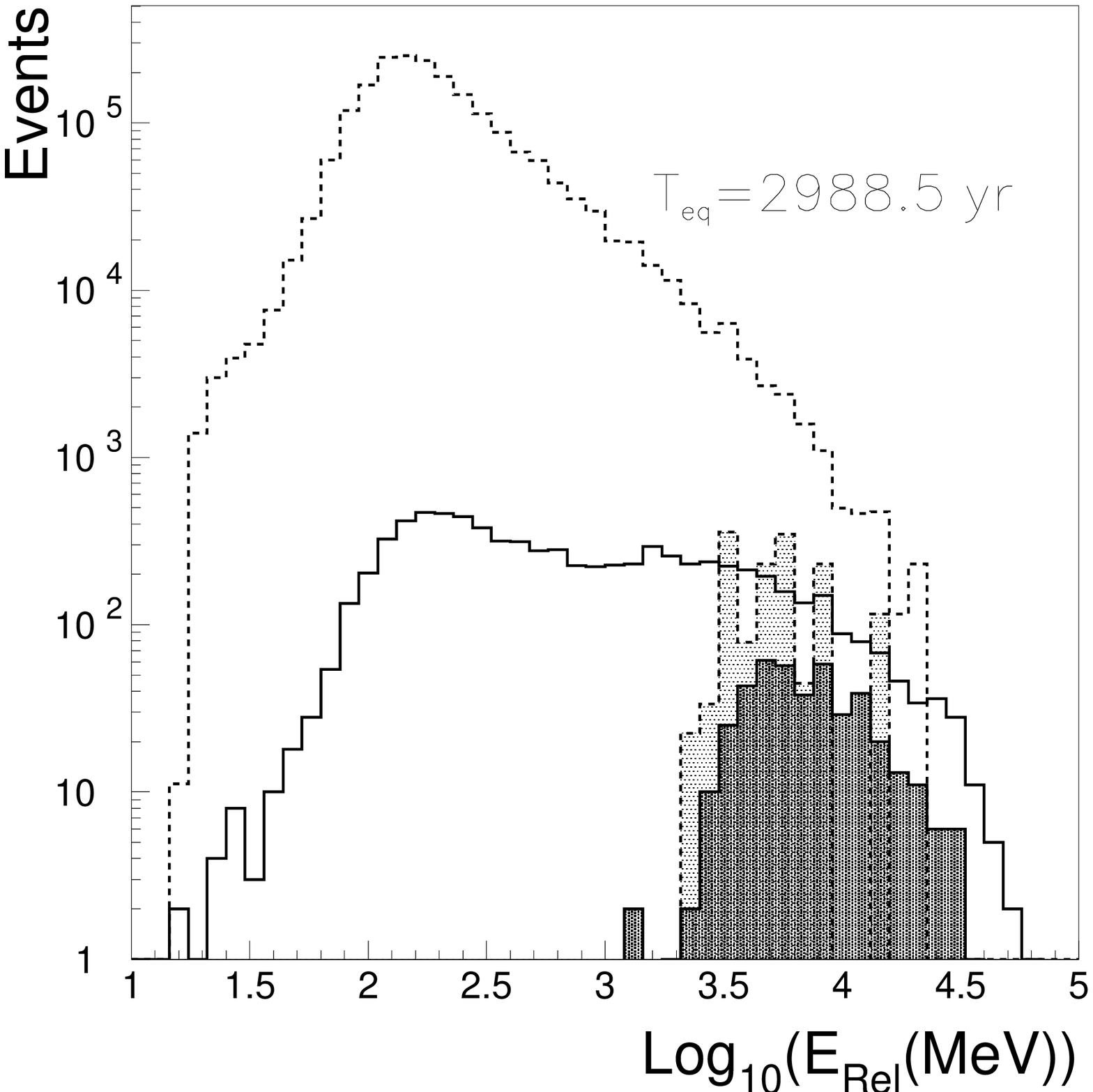,height=7.8cm,width=7.6cm}
\caption{ {\it (Left)}: distribution of the muon initial energy for the AGN 
sample (solid line) and for the atmospheric neutrino background (dashed line).
The events surviving $E_{cut}$ + $D_{cut}$ are shown as the light shaded 
histogram  (atmospheric sample) and as the dark shaded histogram
(AGN sample). {\it (Right)}: distribution of the reconstructed energy released in 
the scintillators, with exactly the same notation as for the plot in the left.}  
\label{eneinirele}
\end{center}
\end{figure}

Fig.~\ref{eneinirele} (left) shows the distribution of the muon initial energy 
(i.e. muon energy when entering the detector) for the atmospheric neutrino 
simulated sample (dashed line) and for the AGN sample (solid line).
The events surviving the analysis cuts $E_{cut}$ + $D_{cut}$ are also shown 
(light shaded histogram for the atmospheric sample, dark shaded histogram for 
the AGN sample).
The Fig.~\ref{eneinirele} (right) shows the distribution of the reconstructed 
energy released in the scintillator counters, with the same notation of the previous 
plot. The two distributions (atmospheric neutrino background and AGN signal) 
have been normalized to the same equivalent time of the AGN sample.
A small number of background events survive the analysis cuts $E_{cut}$ + 
$D_{cut}$. 
Because of the Monte Carlo generation statistics, the error bars (not shown) of 
the light shaded histograms, are at the level of 50\% for each energy bin.

As a final remark, it should be noticed that the simulation of the atmospheric 
neutrino background does not take into account the isotropic contribution of 
neutrinos from semileptonic decays of charmed hadrons, usually called "prompt 
neutrinos".  For neutrino energies above 10 TeV~\cite{prompt1,prompt2,costa} 
this contribution should begin to dominate over the conventional pion and kaon 
decay induced neutrino flux.
Actually, mainly due to the lack of precise information about high energy charm 
inclusive cross sections, the expected fluxes of prompt neutrinos, as well      
as the energy above which they start to dominate, fluctuate on a very wide 
range. 
Even by assuming one of the highest prompt neutrino flux prediction given 
in~\cite{costa}, the contribution to the signal has been estimated as few 
percent of the conventional atmospheric neutrino background (at least in the 
energy range in which this analysis is sensitive).

In Fig.~\ref{uplimfig} the contribution from prompt neutrinos, according to the 
maximum (PN$_{max}$) and the minimum (PN$_{min}$) predictions given in 
\cite{costa}, are shown, together with the flux of conventional atmospheric 
neutrino and the diffuse fluxes of neutrinos from AGNs and GRBs according to  
some predictions.  
%
%
\section{Data analysis and results}
\label{sec::5}
The data used for this analysis were collected in the period from April 1994 to 
December 2000 (5.8 years, including efficiencies).
After imposing the energy cut, $E_{cut}$, and the flight direction cut, 
$D_{cut}$, 97 real events survive. \\
Three events were rejected due to the action of a cut on the quality of the 
data. This quality cut (applied in all MACRO analysis which uses the ERP TDCs) 
requires that the two TDCs, which have a different dynamic range, on each ERP 
channel must agree. 
This inefficiency, calculated from different analysis, affect 2\% of the data 
sample \cite{upmu,lowenergy}.

\begin{figure}[htb]
\begin{center}
\epsfig{file=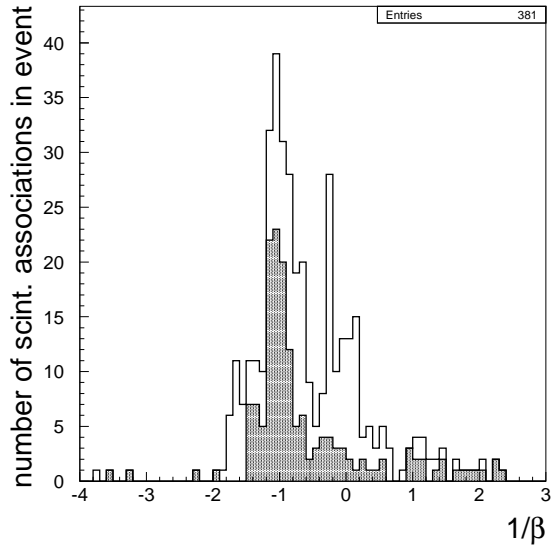,height=8cm,width=8cm}
\caption{$1/\beta$ distribution for one of the two events (Run 16399, Evt 2925) 
surviving all the analysis cuts. One entry corresponds to one $scintillator\ 
track$. 
The $scintillator\ tracks$ fulfilling both the conditions on geometrical length 
 and on timing quality are shown as the shaded 
histogram.}  
\label{piccofig}
\end{center}
\end{figure}

As expected, a large contamination from multiple muons and near horizontal 
events is found. Those background events were rejected with the cut $S_{cut}$; 
after applying this cut, only two events survive.   
Fig.~\ref{piccofig} shows the distribution of the quantity  $1/\beta$ for one of 
the two selected event. The $1/\beta$ for the scintillator tracks fulfilling the 
conditions on geometrical length and on timing quality are shown as the shaded 
histogram.

An attempt to reconstruct the direction of both events has been performed by 
combining scintillator counters with streamer tubes tracking information. Because of the 
strong showering activity, the result is affected by a too high uncertainty for 
any directional astronomy study.     

\begin{figure}[b]
\begin{center}
\epsfig{file=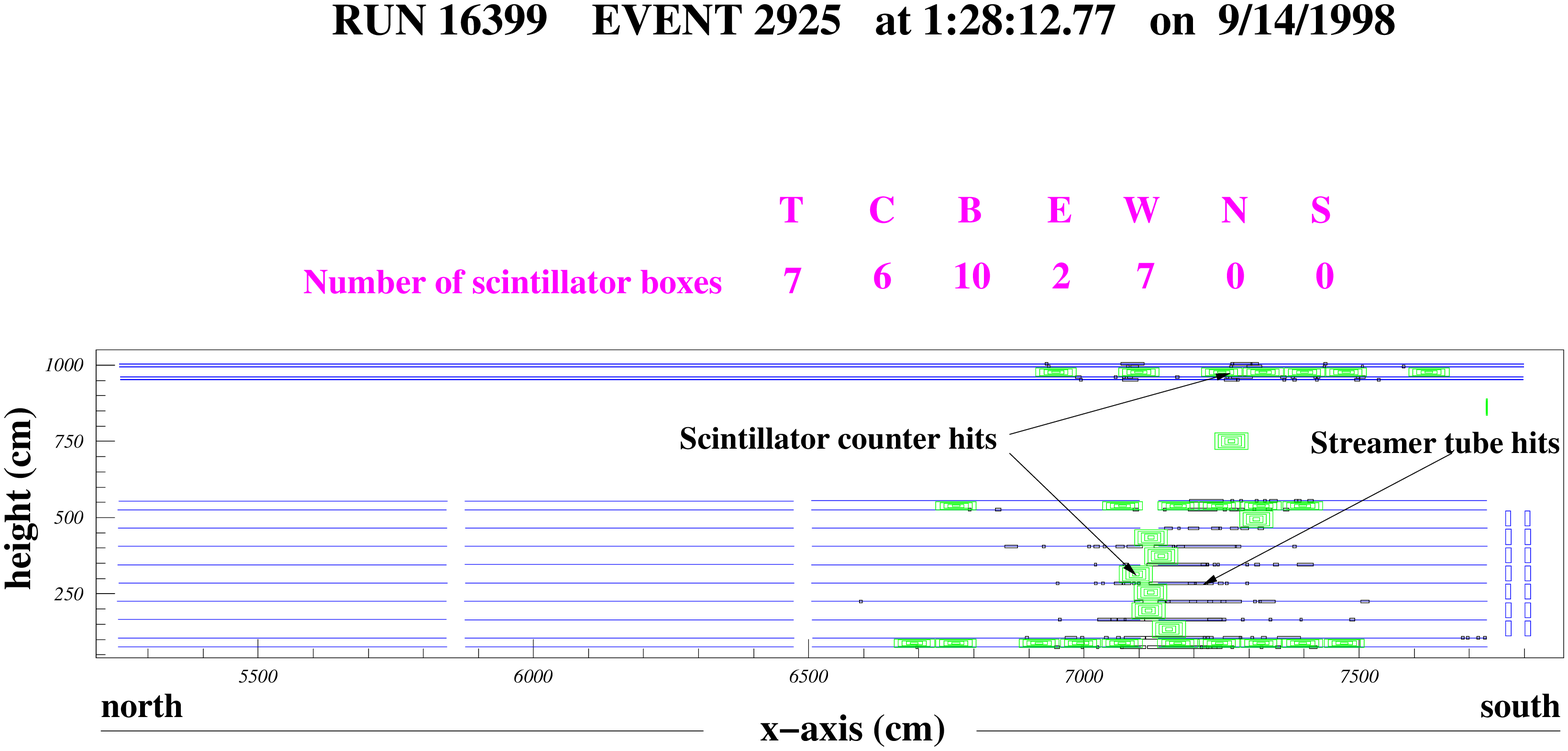,height=11cm,width=13.5cm} 
\hskip -0.8 cm 
\vskip -2. cm
\epsfig{file=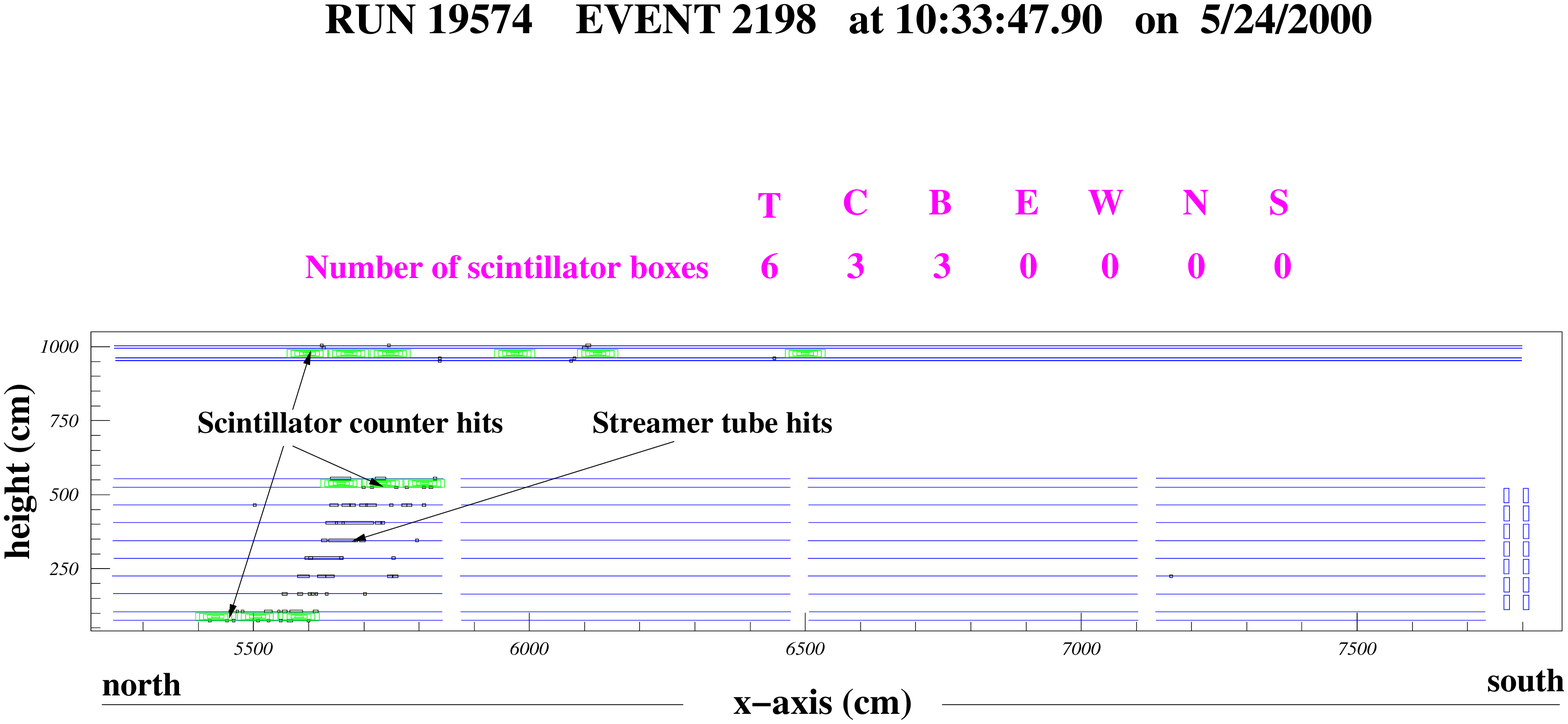,height=11cm,width=13cm} 
\vskip -1.5cm 
\caption{Display of the two surviving events (longitudinal view) along the $x$-axis. 
The horizontal lines represent the 10+4 planes of horizontal streamer tube 
wires in the bottom and in the "Attico" parts of the detector. The wire hits are 
represented by black points.
The gray boxes represent scintillator hits. 
 32 scintillation counters overall fired in 
the first event which correspond to 381 $scintillator\ tracks$, whose $1/\beta$ 
values are shown in Fig. 5. 12 counters fired in the second event.
The location of fired scintillation counters is also given: T$\rightarrow$ Top, C$\rightarrow$ Central,
B$\rightarrow$ Bottom, E$\rightarrow$ East, W$\rightarrow$ West, N$\rightarrow$ North, 
 S$\rightarrow $ South, respectively.}
\label{display}  
\vskip 1.8cm
\end{center}
\end{figure}

The simulated AGN signal events have been processed with the same analysis chain 
as the real data; they are reduced by a factor (2\%) which takes into account 
the above ERP inefficiency, not explicitly reproduced in the simulation of the 
detector response.  
242/13305 simulated upward-going muons from diffuse AGNs survived all 
cuts. These events correspond to 0.54 for 5.8 years live time. 
For the atmospheric neutrino background, 1.1 events are expected in the same 
running period.  The results are summarized in Table~1.
The MACRO display of the two real events is shown in Fig.~\ref{display}. 
\subsection{The upper limit} 
The data show no evidence of possible excess due to a diffuse neutrino flux from 
AGNs within the framework of the model \cite{Stecker}.  
Our result was used to set a muon flux upper limit $F_{\mu}^{L}$ which was 
calculated at 90$\%$ confidence level:  
\begin{equation}
 F_{\mu}^{L}= \frac{{\rm Upper \; limit} 
(90\% c.l.)} {\epsilon \times \int Area(\Omega) d\Omega  
    \times T_{l}}
\label{upperlimit}
\end{equation}   
The calculation of the numerator is based on the observation of 2 events 
when the 
expected background is 1.1 events, according to the recent approach of Feldman 
and Cousins~\cite{Feldman}.
$\epsilon$ is the fraction of simulated AGN events which survive the analysis 
cuts; $Area(\Omega)$ is the geometrical area seen by the diffuse flux as a 
function of the solid angle $\Omega$ (the integral extends to the lower Earth 
hemisphere) and $T_{l}=5.8$ years is the considered MACRO live time.  
With these values the muon flux upper limit becomes: 
\begin{equation}
\boxed{F_{\mu}^{L}(90\% \, c.l.)= (1.7 \pm 0.2) \times 10^{-14} 
 \mathrm{cm^{-2}} \, \mathrm{s^{-1}}
 \, \mathrm{sr^{-1}}}
\label{uplim}                  
\end{equation}
This limit can be finally converted into a (differential) neutrino flux upper 
limit. In order to compare our results with the upper limits given by other 
experiments, it is convenient to assume a power law spectrum with spectral index 
2 for the initial neutrino flux.       
We have then weighted the number of surviving events with the ratio between this 
spectrum and the spectrum  calculated by Stecker {\it et al.}.  
With these hypotheses, the neutrino flux upper limit from this analysis is 
$E^{2} \cdot F_{\nu}^{L}= $ $(4.1 \pm 0.4)$ $\times 10^{-6} \,  
\mathrm{GeV} \, \mathrm{cm}^{-2} \, \mathrm{s}^{-1}
 \, \mathrm{sr}^{-1}$.
Actually, this value must be considered carefully because it is affected by 
model assumptions (initial spectrum, neutrino cross sections and muon energy 
losses).  
Here, it has been estimated for comparison purposes. 
\begin{table}
\begin{center}
\begin{tabular}{||c|c||}\hline\hline
  &  Rate of survived events in 5.8 y\\ \hline
{\bf \it Atmospheric (MC)} & 1.1 $\pm 0.5$ \\
{\bf \it Agn (MC)} & 0.54 $\pm 0.03$  \\
 {\it DATA} & 2\\
 \hline \hline
\end{tabular}
\\[2pt]
\vskip 0.4cm  
\end{center}
Table 1: Expected rate of events surviving all analysis cuts in 5.8 years of 
MACRO running. 
Besides the number of events in the data, the expectation from the 
atmospheric background and the AGN neutrino flux (according to~\cite{Stecker}) 
are shown. Only the statistical errors of the simulation were considered.
\end{table}

\begin{figure}[bt]
\begin{center}
\epsfig{file=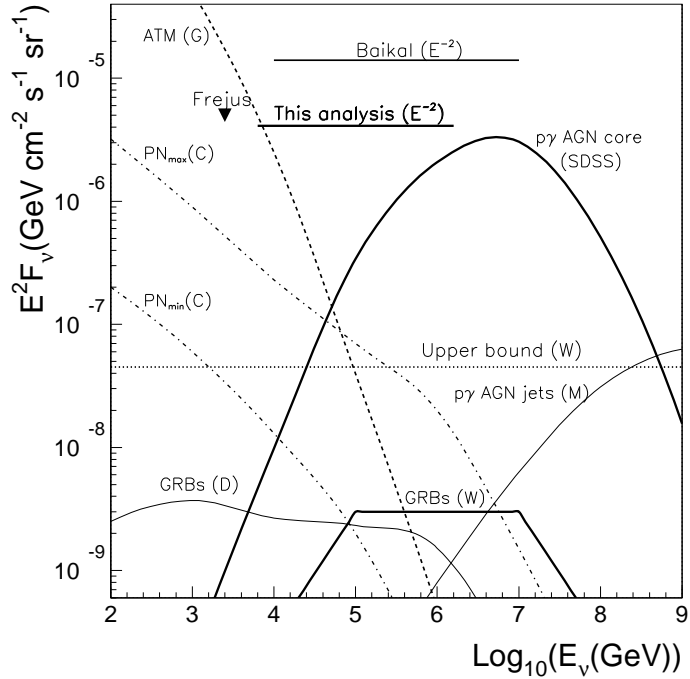,height=10cm,width=10cm}
\caption{
Diffuse fluxes of $\nu_{\mu}+\bar{\nu}_{\mu}$ from AGNs and GRBs
 according to many predictions. (SDSS) Stecker {\it et al.},
1991~\protect\cite{Stecker}; (M) Mannheim, 1995~\protect\cite{mannheim};
  (W) Waxman and Bahcall, 1998~\protect\cite{waxman}; 
  (D) De Paolis {\it et al.}, 2001~\protect\cite{nugrb}. 
  The dashed curve refers to the angle average atmospheric
  neutrino flux; (G) Gaisser {\it et al.},1995~\protect\cite{gp}. 
    The dash-dotted lines refer to the flux of prompt neutrinos 
   according to the maximum (PN$_{max}$) and the  
   minimum (PN$_{min}$) predictions given in Costa, 2001
\protect\cite{costa}.
  The dotted line is the theoretical upper bound to neutrino 
   fluxes from astrophysical sources as calculated  
  in~\protect\cite{waxman}. 
Some current published experimental upper limits are also
   shown. See Tab. 2 for references to the experiment results.}  
\label{uplimfig}
\end{center}
\end{figure}

Fig.~\ref{uplimfig} shows some theoretical predictions  of neutrino fluxes from 
astrophysical sources (AGNs and GRBs) in comparison with upper limits obtained 
by current experiments. A list of them with references and energy range of 
sensitivity  is given in Table~2.      
\begin{table}[htb]
\begin{center}
\begin{tabular}{||c|c|c||}\hline\hline
  &  $E^{2} \cdot F_{\nu}^{L}$ (90\% C.L.)  
   & Neutrino Energy Range \\ 
&  ($\mathrm{GeV} \, \mathrm{cm}^{-2} \, \mathrm{s}^{-1}
 \, \mathrm{sr}^{-1}$) & (GeV)\\
 \hline
 EAS-TOP~\cite{eastop} & 2.0 $\times 10^{-3}$ & $10^{5} \div 10^{6}$\\
SPS-DUMAND~\cite{dumand} & 6.0 $\times 10^{-4}$ & $10^{5} \div 10^{6}$ \\
Baikal~\cite{baikal} & 1.4 $\times 10^{-5}$ & $10^{4} \div 10^{7}$ \\
Baikal ($\nu_{e}$)~\cite{baikalnue} & 1.3 $\div$ 1.9 $\times 10^{-6}$ &
$10^{4} \div 10^{7}$ \\
Frejus~\cite{Frejus} & 5.0 $\times 10^{-6}$ &  $\sim$ 2.6$\times 10^{3}$ \\
MACRO (this analysis) & 4.1 $\pm$ 0.4 $\times 10^{-6}$ & $10^{4} \div 10^{6}$ \\
AMANDA~\cite{amandadiff}& 1.0 $\times 10^{-6}$ & $10^{3} \div 10^{6}$  \\
 \hline \hline
\end{tabular}
\\[2pt]
\vskip 0.4cm  
\end{center}
Table 2: Neutrino flux upper limits (90\% C.L.) from current experiments. 
 The neutrino sensitivity range is shown in the last column.
\end{table}

%
\section{Conclusions}
We presented the results of a search for a diffuse neutrino flux from unresolved 
astrophysical sources by analyzing a sample of high energy events collected by 
MACRO.  
We tested the detector response at high energy taking special care for the 
reliability of the scintillation counter response to high-energy particles.  We 
used the reconstructed energy released in the scintillators to select a sample 
of high energy events. 
Two high energy upward-going candidates have been found in 5.8 y of effective 
running time. This number is compatible with the expected atmospheric neutrino 
background (1.1 events). 
The analyzed data have been used to set a muon and neutrino flux upper limit 
comparable with the results given by other experiments.

\section{Acknowledgments}

We gratefully acknowledge the support of the director and of the staff of the 
Laboratori Nazionali del Gran Sasso and the invaluable assistance of the 
technical staff of the Institutions participating in the experiment. We thank 
the Istituto Nazionale di Fisica Nucleare (INFN), the U.S. Department of Energy 
and the U.S.  National Science Foundation for their generous support of the 
MACRO experiment. We thank INFN, ICTP (Trieste), WorldLab and NATO for providing 
fellowships and grants (FAI) for non Italian citizens.

\newpage

\end{document}